\documentclass[article,prc,aps,epsfig,floatfix,twocolumn]{revtex4}
\usepackage{amssymb}
\usepackage{graphicx,psfrag}

\newcommand{\bld}[1]{\mbox{\boldmath $#1$}}	
\newcommand{\code}{{\tt FREYA}}			
\newcommand{\ie}{{\em i.e.}}			
\newcommand{\etal}{{\em et al.}}		

\begin{document}

\title{Study of angular momentum effects in fission}

\author{R.~Vogt$^{1,2}$ and J.~Randrup$^3$}

\affiliation{
  $^1$Nuclear and Chemical Sciences Division,
  Lawrence Livermore National Laboratory, Livermore, CA 94551, USA\break
  $^2$Physics and Astronomy Department, University of California,
  Davis, CA 95616, USA\break
$^3$Nuclear Science Division, Lawrence Berkeley National Laboratory, 
Berkeley, CA 94720, USA}

\date{October 10, 2020}

\begin{abstract}
{\bf Background:}  The role of angular momentum in fission has long been
discussed but the observable effects are difficult to quantify.  
  {\bf Purpose:}  We discuss a variety of effects associated with angular
  momentum in fission and present quantitative illustrations.
  {\bf Methods:}  We employ the fission simulation model \code\ which is well
  suited for this purpose because it obeys all conservation laws, including
  linear and angular momentum conservation at each step of the process.  We
  first discuss the implementation of angular momentum in \code\ and then
  assess particular observables, including various correlated observables.
  We also study potential effects of neutron-induced fission of the low-lying
  isomeric state of $^{235}$U relative to the ground state.
  {\bf Results:}   The fluctuations inherent in the fission process 
  ensure that the spin of the initial compound nucleus has only a small
  influence on the fragment spins which are therefore nearly uncorrelated.
  There is a marked correlation between the spin magnitude
  of the fission fragments and the photon multiplicity.  We also consider
  the dynamical anisotropy caused by the rotation of an evaporating fragment
  and study especially the distribution of the projected neutron-neutron
  opening angles, showing that while it is dominated by the effect of the
  evaporation recoils,
  it is possible to extract the signal of the dynamical anisotropy by means of
  a Fourier decomposition.  Finally, we note that the use of an isomeric target,
  $^{235{\rm m}}$U($n_{\rm th}$,f), may enhance the symmetric yields and can
  thus result in higher neutron multiplicities for low total fragment kinetic
  energy.

\end{abstract}

\maketitle

\section{Introduction}

The role of angular momentum in nuclear fission has long been
a topic of central interest, dating back over sixty years.
Experimental evidence suggests that the primary fission fragments
on average carry spins of magnitude $5-7 \hbar$
aligned roughly perpendicular to the fission axis \cite{Wilhelmy}.
The associated fragment rotation generally causes the neutron evaporation
to be anisotropic \cite{Bowman,Gavron}
which may affect a variety of neutron-related observables,
including neutron spectra, angular distributions,
and directional correlations \cite{Pringle},
as well as attempts to find evidence of scission neutrons
\cite{Bowman,Vorobyev,Chietera}.
The fragment angular momentum also influences the photon radiation
\cite{VR-PRC96}.

For the purpose of elucidating the various ways angular momentum enters
into the fission process and to quantitatively ascertain
the effect on observables of particular interest,
the present study utilizes the event-by-event fission model
\code \cite{CPC191,CPC222}, which uses Monte Carlo techniques 
to generate large samples of complete fission events.
An important advantage of employing \code\ is that all conservation laws
are obeyed throughout each fission event, including those 
affecting the angular momentum directions during the evaporation cascades.

Section \ref{FREYA} describes how angular momentum is treated in \code.
Section~\ref{neutrons} addresses observables of particular interest.
Then Sec.~\ref{isomer} describes potential observable differences in
$^{235}$U($n_{\rm th}$,f) arising if the target nucleus is in its isomeric state.
Our concluding remarks are presented in Sec.~\ref{summ}.

\section{Treatment of angular momentum in FREYA}
\label{FREYA}

\code\ is a Monte Carlo model that is capable of quickly generating 
large samples of complete fission events,
namely the full kinematic information for the two prompt product nuclei
and all prompt neutrons and photons in each event.
With these large samples, it is straightforward to extract
any observable of interest.
Because each fission event conserves mass, charge, energy, linear and
angular momentum, and spin, 
any inherent correlations between various quantities
are preserved, thus making \code\ particularly well suited
for determining how angular momentum affects various final states.

Angular momentum enters at several stages during the fission process.
In the following section, we describe how it is treated in \code.

\subsection{Preparation}

During the first stage of a neutron-induced fission event, 
a neutron of momentum $\bld{p}_0$ impinges on the target nucleus,
in the present case $^{235}$U.
The associated impact parameter \bld{\rho}
is chosen randomly from a disk, perpendicular 
to the direction of motion, with a radius equal to that of the nucleus.
Thus the neutron introduces a linear momentum of $\bld{p}_0$
and an angular momentum of $\bld{\rho}\times\bld{p}_0$.

The incoming neutron may cause the emission of a pre-equilibrium neutron.
While pre-equilibrium processes grow increasingly likely at higher energies,
they are unimportant at thermal energies.
If pre-equilibrium emission does occur, the appropriate reduction is made
of the mass number as well as the linear and angular momenta
of the residual system which is assumed to 
subsequently relax to a compound nucleus which can ultimately fission.

If sufficiently excited, the compound nucleus may evaporate
one or more neutrons before fission occurs, according to 
the energy-dependent branching ratio $\Gamma_{n}/\Gamma_{\rm f}$.
Such pre-fission evaporation is treated the same way as the
post-fission evaporation from the fission fragments, see Sec.~\ref{decay}.
For each evaporation, \code\ reduces the excitation energy 
of the daughter nucleus and changes its linear and angular momenta
as dictated by conservation laws.

At the end of the pre-fission evaporation chain,
we arrive at the fissioning nucleus with angular momentum $\bld{S}_0$
which has generally been reoriented relative to $\bld{\rho}\times\bld{p}_0$
due to the spin recoils, by about $24^\circ$ on average
for $^{235}$U(n,f) for an incoming neutron energy of $E_n=20$~MeV.

\subsection{Scission}
\label{sciss}

The second stage of the fission process is the evolution of the
pre-fission compound nucleus to two well-separated 
and fully-accelerated primary fragments,
each of which is also in equilibrium.
It is not the purpose of \code\ to model how this complicated
time-dependent many-body process develops.
[That challenging problem was recently reviewed in Ref.\ \cite{FoF}
and a broader review of the recent experimental and theoretical progress
in fission can be found in Ref.\ \cite{SchmidtJurado}.]
Rather, \code\ generates an ensemble of possible outcomes based 
primarily on the input provided.
Generally, the \code\ input is based on experimental data,
but a certain degree of modeling is necessary 
because the data sets are often incomplete.
For example, while experimental data for the fragment mass distribution $Y(A)$ 
and the associated mean total fragment kinetic energy $\overline{\rm TKE}(A)$
is often available for thermal neutron-induced fission,
the energy dependence of these functions is usually not well measured.

After the neutron and proton numbers of the primary fragments have been 
selected, their angular momenta are sampled from the statistical distribution 
of the dinuclear rotational modes at scission \cite{RV-PRC89}.
The total angular momentum of the system after scission
is given by $\bld{S}_0=\bld{S}_L + \bld{S}_H + \bld{L}$,
namely the sum of the two individual light and heavy fragment spins and the
orbital angular momentum of the two-fragment system,
$\bld{L}=\bld{R}\times\bld{P}$,
where $\bld{R}=\bld{R}_L - \bld{R}_H$ is the fragment separation
and $\bld{P}=\mu(\bld{V}_L - \bld{V}_H)$ is the associated momentum,
equal to the reduced mass 
$\mu=M_L M_H/(M_L + M_H)$ times their relative velocity.
The overall rotation of the dinuclear complex
determines the average values of the fragment spins,
$\overline{\bld{S}}_i=({\cal I}_i/{\cal I})\bld{S}_0$, 
as well as the average of their relative angular momentum, 
$\overline{\bld{L}}=({\cal I}_R/{\cal I})\bld{S}_0$.
Here ${\cal I}_i$ is the moment of inertia of fragment $i = L,H$,
${\cal I}_R=\mu R^2$ is the moment of inertia for the relative motion, and
${\cal I}={\cal I}_L + {\cal I}_H + {\cal I}_R$ is the total moment of inertia.
(The nuclear moments of inertia are taken 
to be 50\%\ of the rigid-body values, as is commonly done.)

Because the system is excited at the time of scission,
the intrinsic rotational modes are expected to be agitated 
and the actual angular momenta will therefore fluctuate,
$\bld{S}_i=\overline{\bld{S}}_i+\delta\bld{S}_i$.
In order to sample the spin fluctuations, $\delta\bld{S}_i$,
\code\ brings the rotational energy,
\begin{equation}
E^{\rm rot} = S_L^2/2{\cal I}_L + S_H^2/2{\cal I}_H + L^2/2{\cal I}_R\ ,
\end{equation}
onto normal form \cite{CPC222,RV-PRC89}.
A binary system generally has six normal modes of rotation \cite{DR-NPA433}. 
These are {\em tilting} and {\em twisting}, 
in which the fragments rotate in the same or in the opposite sense 
around the dinuclear axis $\hat{\bld{z}}=\bld{R}/R$,
and {\em wriggling} and {\em bending},
in which they rotate in the same or in the opposite sense around 
an axis perpendicular to the dinuclear axis \cite{NixNP71,MorettoPRC21}.
The two latter modes are each doubly degenerate (corresponding to 
rotations around $\hat{\bld{x}}$ and $\hat{\bld{y}}$, for example.)
Only the perpendicular modes (wriggling and bending)
are considered by \code\ \cite{CPC222,RV-PRC89},
because the agitation of the first two tends to be suppressed 
due to the constricted neck \cite{DR-NPA433}.

The rotational energy associated with the four perpendicular modes
can be written on normal form as
\begin{equation}
E_\perp^{\rm rot} = s_+^2/2{\cal I}_+ + s_-^2/2{\cal I}_-\ ,
\end{equation}
where $\bld{s}_+$ represents wriggling
and $\bld{s}_-$ represents bending.
The corresponding moments of inertia are \cite{DR-NPA433}
\begin{equation}
{\cal I}_+=({\cal I}_L + {\cal I}_H){\cal I}/{\cal I}_R\ ,\,\,\
{\cal I}_-={\cal I}_L {\cal I}_H/({\cal I}_L + {\cal I}_H)\ .\end{equation}
\code\ then samples $\bld{s}_\pm$ from a distribution of statistical form,
$P(\bld{s}_\pm)\sim\exp(-s_\pm^2/2{\cal I}_\pm T_S)$,
where $T_S$ is the effective spin temperature (explained below).

The resulting angular momenta of the individual fragments 
are subsequently obtained as
\begin{eqnarray}\label{S1}
\bld{S}_L\!\!&=&\! \overline{\bld{S}}_L + \delta\bld{S}_L
=	  ({\cal I}_L/{\cal I})\bld{S}_0
	  +({\cal I}_L/{\cal I}_+)\bld{s}_+ +\bld{s}_- ,\,\,\\ \label{S2}
\bld{S}_H\!\!&=&\! \overline{\bld{S}}_H + \delta\bld{S}_H\!
=	  ({\cal I}_H/{\cal I})\bld{S}_0\!
	+({\cal I}_H/{\cal I}_+)\bld{s}_+\!-\bld{s}_- .
\end{eqnarray}
The fluctuations $\bld{s}_+$ and $\bld{s}_-$ are oriented 
randomly in the plane perpendicular to the dinuclear axis.
The wriggling mode adds parallel fluctuations to the fragment spins,
while the contributions from the bending mode are anti-parallel.
There is thus no simple relationship between the direction
of the two resulting fragment spins, $\bld{S}_L$ and $\bld{S}_H$,
as we now discuss in more detail.

As brought out in Eqs.~(\ref{S1})-(\ref{S2}),
there are two distinct contributions to the angular momentum
of a primary fragment, namely the fragment's share of the overall rotation 
of the dinuclear complex, $\overline{\bld{S}}_i$,
and the fluctuations received at scission, $\delta\bld{S}_i$.
It is important to recognize that the latter generally dominate.

To understand this important feature, we first note that 
the angular momentum brought in by a thermal neutron is negligible, 
$S_0\approx 0.34\hbar$ on average.
At an incoming energy of $E_n=20$~MeV 
the angular momentum of the fissioning compound nucleus is $S_0\approx 5\hbar$.
However, at scission where the total angular momentum is divided up
between the two fledgling fragments and their relative motion,
the former acquire only small fractions because their moments of inertia 
are relatively small, ${\cal I}_i\ll{\cal I}_R$.
Consequently, even for $E_n=20$~MeV, 
this contribution amounts to only $\approx\!0.26\hbar$ on average
for $^{235}$U(n,f)
(this would increase to $\approx\!0.48\hbar$ if the full rigid values
were used for the fragment moments of inertia).

The magnitude of the spin fluctuations is governed by the
degree of internal excitation at scission, $E_{\rm sc}^*$.
Because this quantity depends on how much of the total excitation energy, TXE,
is tied up in distortion energy, it is not readily available
and \code\ therefore employs an effective value given by
$c_S^2\!\cdot\!{\rm TXE}=a_0T_S^2$,
where the reduction factor $c_S$ is a parameter in \code\
and the level-density parameter $a_0$ is calculated 
assuming a back-shifted Fermi gas \cite{RV-PRC80}.
[For $^{235}$U(n,f) \code\ uses $c_S=0.87$;
one can elucidate the effects of the spin fluctuations 
by varying $c_S$, see Ref.~\cite{VR-PRC96}.]
Starting from $\approx\!22$~MeV for thermal fission,
TXE increases steadily to $\approx\!40$~MeV at $E_{n}=20$~MeV,
so the effective spin temperatures $T_S$ is 
in the range $0.85 - 1.15$~MeV.
The mean spin fluctuations $[\langle(\delta S_i)^2\rangle]^{1/2}$
are then $4.8\hbar/6.4\hbar$, for the light/heavy fragment in thermal fission
and $5.8\hbar/7.1\hbar$ for $E_n=20$~MeV
(or about two units more if the rigid body moments of inertia were used).
Thus, generally, the spin fluctuation is over an order of magnitude larger 
than the aligned component. Consequently, the fragment spins are
primarily determined by the fluctuations acquired at scission.

As discussed above, the fragments themselves inherit only a small fraction 
of the total angular momentum in the system, 
with the main part going to the relative motion.  
Because the spin fluctuations dominate over the averages,
there is very little correlation remaining between the directions of the 
resulting total fragment spins $\bld{S}_i$ 
and the direction of the overall angular momentum, $\bld{S}_0$, 
aside from them all being perpendicular to the dinuclear axis.
(Note that the relative orbital angular momentum $\bld{L}$ is adjusted 
to counteract the bending-mode fluctuations,
ensuring conservation of the total angular momentum.)

Furthermore, the two individual fragment spins are also rather uncorrelated.
Indeed, ignoring the small aligned component $\overline{\bld{S}}_i$
(see Eqs.\ (\ref{S1}) and (\ref{S2})),
\ie\ assuming $\bld{S}_i\approx\delta\bld{S}_i$, we find
\begin{eqnarray}\label{Si2}\nonumber
& \langle{S}_i^2\rangle \approx
  ({\cal I}_i^2/{\cal I}_+^2) \langle{s}_+^2\rangle
+ \langle{s}_-^2\rangle
=2\left(({\cal I}_i^2/{\cal I}_+) +{\cal I}_-\right)T_S\\
& = 2{\cal I}_i \left(1-{\cal I}_i/{\cal I}\right)T_S
\approx 2{\cal I}_iT_S\ .\phantom{nmmmmm}
\end{eqnarray}
Thus the resulting fragment spins are approximately equivalent to 
statistical sampling without preserving any conservation laws.
The ratio between the wriggling and bending terms in Eq.~(\ref{Si2}) is
$\approx{\cal I}_{H}/{\cal I}_{ L}$ 
	      for $\langle{S}_{ L}^2\rangle$, while it is
$\approx{\cal I}_{ L}/{\cal I}_{ H}$
    	      for $\langle{S}_{ H}^2\rangle$.
Thus the two types of modes contribute about equally to the fragment spins.
It then follows that their directional correlation is rather weak.

\begin{figure}[tbh]	   
\includegraphics[width=0.95\columnwidth]{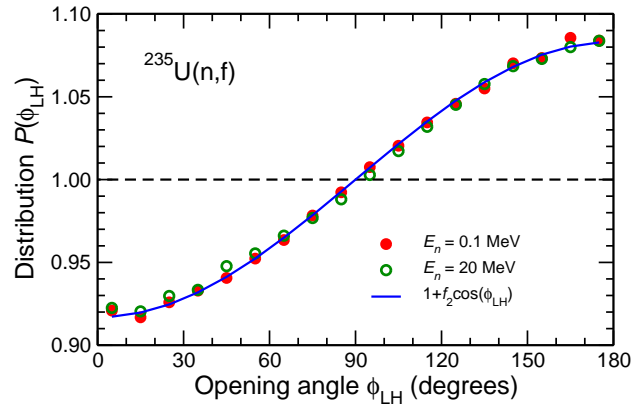}
\caption{(Color online) 
The distribution of the opening angle $\phi_{LH}$ 
between the spins of the two primary fission fragments from $^{235}$U($n$,f)
for two different neutron energies: thermal neutrons and $E_{n}=20$~MeV.
Also shown is the lowest-order Fourier approximation, 
$P(\phi_{LH}) = 1+f_2 \cos\phi_{LH}$.
}\label{fig1}
\end{figure}

This overall weak directional correlation can be quantitatively observed
in Fig.~\ref{fig1} 
which shows $P(\phi_{LH})$,
the distribution of the opening angle between the two spins.
For a given mass partition,
this distribution does not depend on the incident energy
because the spin fluctuations scale with $T_S$.
Furthermore, because the moments of inertia of the fragments
are so small compared to the relative moment of inertia
there is also very little dependence on the mass partition.
Thus $P(\phi_{LH})$ is a fairly universal function.
We note that it can be represented to a very good approximation as
$P(\phi_{LH})\approx1+f_2\cos\phi_{LH}$,
with Fourier amplitude $f_2\approx-0.082$.  Thus
the two fragment spins have a slight preference for being
oppositely directed, with $P(180^\circ)/P(0^\circ) \approx 1.18$.

It should be noted that because the wriggling mode contributes 
parallel spin fluctuations,
conservation of angular momentum causes 
the orbital angular momentum \bld{L}\ to be affected oppositely,
$\delta\bld{L} = -({\cal I}_R/{\cal I})\bld{s}_+$.
This changes not only the magnitude of \bld{L}\ but also its orientation.
Therefore the plane of the relative fragment motion (the exit plane)
generally differs from the impact plane.
\code\ takes this into account when calculating the orbital Coulomb trajectory 
of the receding fragments.
Furthermore, because the Coulomb trajectory is hyperbolic,
the asymptotic direction of the relative fragment motion
differs from the orientation of the system at scission.
However, as a result of the relative slowness of the orbital fragment motion 
at scission and the strong radial acceleration from the mutual Coulomb
repulsion, the associated reorientation angle is rather small,
amounting typically to about $2^\circ$.

\subsection{Fragment de-excitation}
\label{decay}

After their formation and acceleration,
the excited primary fragments undergo a sequence of decay processes.
\code\ first considers neutron evaporation,
starting by calculating the available statistical excitation energy for each
fragment,
$Q_i=E_i^*-E_i^{\rm rot}$,
where $E_i^*$ is its total excitation energy
and its rotational energy is given by $E_i^{\rm rot}=S_i^2/2{\cal I}_i$.
If the statistical energy exceeds the neutron separation energy, $S_{n}$,
then evaporation can occur.
The neutron is evaporated with a black-body spectrum
from a randomly selected point on the surface of the rotating fragment.
The local rotational velocity of the surface element 
adds a centrifugal boost to the neutron.
The daughter fragment absorbs the resulting linear and angular momentum recoils.
This procedure is iterated as long as evaporation is energetically allowed.

\begin{figure}[tbh]    
\includegraphics[width=0.95\columnwidth]{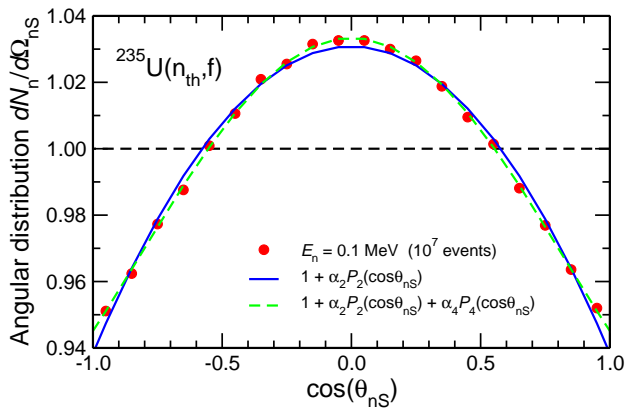}
\caption{(Color online) 
The angular distribution of the evaporated neutrons relative to 
the spin direction of the emitting fragment, $dN_{n}/d\Omega_{ nS}$,
averaged over the evaporation chains for 
the entire fragment yield distribution from $^{235}$U($n_{\rm th}$,f).
In addition to the \code\ simulation results (solid red circles),
the lowest two Legendre approximations are also shown.
}\label{fig2}
\end{figure}

The centrifugal boost from the fragment rotation 
causes the angular distribution of the evaporated neutrons
to be anisotropic, with an enhancement in the equatorial plane,
as illustrated in Fig.~\ref{fig2}.
The degree of bulging may be expressed in terms of 
the so-called dynamical anisotropy \cite{Chietera},
\begin{equation}
\label{A}	A\ \equiv\ 
\left[{dN_{n}\over d\Omega_{nS}}\right]_{\theta_{nS}=90^\circ}
/ \left[{dN_{n}\over d\Omega_{nS}}\right]_{\theta_{nS}=0^\circ} -1\ ,
\end{equation}
which is $\approx\!0.093$ for $^{235}$U($n_{\rm th}$,f).
The fragment evaporation chains lead to a reorientation of the fragment spins
by $\approx\!13^\circ$ on average, while the spin magnitudes are reduced
only very slightly, by $\approx\!0.06\,\hbar$ on average.

After evaporation has ceased,
the resulting product nucleus disposes of its remaining excitation energy
and angular momentum by photon emission.
First the statistical excitation energy is radiated away
by emission of E1 and M1 photons,
each one changing the nuclear spin by one unit.
The statistical radiation brings the system to the yrast line,
where $E_i^*=E_i^{\rm rot}$. The nucleus then 
starts emitting stretched quadrupole photons.
At some point, the excitation reaches the regime
tabulated in the RIPL database \cite{RIPL} and \code\ then simulates
those transitions until the ground state, 
or a sufficiently long-lived isomeric state, is reached.
(The further fate of a prompt product nucleus
due to $\beta$ processes is not yet considered in \code.)

\begin{figure}[t]	   %
\includegraphics[width=0.95\columnwidth]{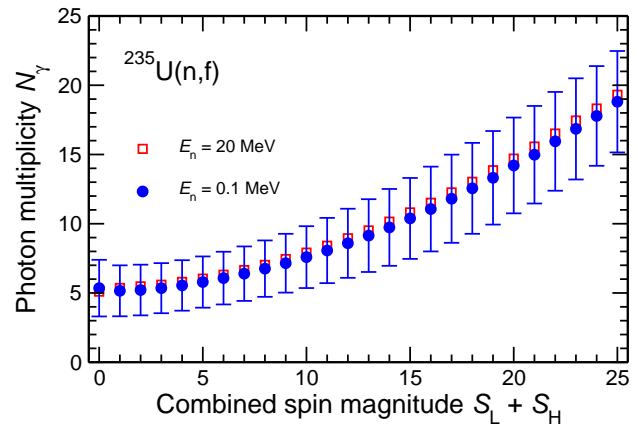}
\caption{(Color online) 
The mean number of photons emitted from $^{235}$U($n_{\rm th}$,f)
(solid blue circles) as a function of the sum of 
the two primary fragment spin magnitudes, $S=S_L+S_H$.
The associated event-by-event photon multiplicity dispersion 
is indicated by the vertical bars.
Also shown are the mean multiplicities for $E_{n}=20$~MeV 
(open red squares).
}\label{fig3}
\end{figure}

For the present study, it is interesting to note that 
there is a relatively tight correlation
between the initial fragment spin magnitude and the number of photons emitted,
as brought out in Fig.~\ref{fig3}. 
This relationship is fairly universal:
it is approximately independent of the incident neutron
energy and it changes by only a fraction of a unit
between $^{233}$U($n$,f) and Cf(sf).
When the combined spin magnitudes, $S\equiv S_{\rm L}+S_{\rm H}$,
exceeds $\approx 6\,\hbar$,
there is a clear increase in $\overline{N}_\gamma$ with $S$,
with roughly one additional photon emitted
for each additional unit of total fragment angular momentum.
As may be expected, the relationship is sensitive to the degree of reduction
of the fragment moment of inertia, $c_I\equiv{\cal I}/{\cal I}_{\rm rigid}$.
For $c_I=0.5$, the value used throughout the present study,
the slope for large $S$ is $d\overline{N}_\gamma/dS\approx 0.84$,
while it is $\approx\! 1.06$ for $c_I=0.3$.
Because of this feature, the measurement of the photon multiplicity may, 
to some degree, substitute for the measurement of the total
fragment angular momenta (see Fig.~\ref{fig6}).

\section{Results}
\label{neutrons}

In this section, results are presented for observables that may
elucidate behavior caused by the fragment angular momenta.  We begin with
observables related to the neutron distribution relative to the direction of
the primary fragment.  We then discuss neutron-neutron correlations gated on
the fragment angular momentum, using photon multiplicity as a proxy.
Finally, we examine a recently proposed observable \cite{Chietera}
based on the transverse neutron motion.

As mentioned in the previous section, the neutrons evaporated from 
rotating fission fragments have a slight preference for emission
perpendicular to the angular momentum due to the centrifugal boost.
Simulations with \code\ show that the angular distribution,
\begin{equation}\label{dNdthS}
{dN_{n}\over d\Omega_{nS}} \sim\ 1+
\alpha_2 P_2(\cos\theta_{ nS}) +\alpha_4 P_4(\cos\theta_{ nS}) +\dots,
\end{equation}
where the polar angle $\theta_{ nS}$, defined with respect to 
the direction of the mother fragment spin,
is well described by the second-order Legendre approximation
(see Fig.~\ref{fig2}).
When averaging over all events
(\ie\ over the impact parameter, mass, charge, and TKE,
as well as the associated evaporation cascades),
\code\ gives $\alpha_2 = -0.061$ and $\alpha_4 = 0.0056$
for $^{235}$U($n_{\rm th}$,f).

Because the orientation $\chi$ of the fragment spin in the plane
perpendicular to the fragment motion is unknown,
averaging over $\chi$ turns the inherently oblate emission pattern 
in Eq.~(\ref{dNdthS}) into a prolate shape 
with its symmetry axis along the fragment direction.
In the frame of the moving fragment,
the distribution of $\theta_{ nF}$,
the angle between the neutron velocity and that of the fragment, is 
\begin{equation}\label{dNdthF}
\langle{dN_{n}\over d\Omega_{nF}}\rangle_\chi
\sim 1-\alpha_2'P_2(\cos\theta_{nF})
+\alpha_4'P_4(\cos\theta_{nF})+\dots\ .
\end{equation}
It can be generally shown that the coefficients in Eq.~(\ref{dNdthF})
are related to those in Eq.~(\ref{dNdthS}) by
$\alpha_{2n}'=(-1)^n\langle\cos^{2n}\!\chi\rangle\alpha_{2n}$,
where $\langle\cos^{2n}\!\chi\rangle=(2n!)/[2^{2n}(n!)^2]$ 
is the average over the spin direction $\chi$.
We thus have $\alpha_2'=-\mbox{$1\over2$}\alpha_2$, 
$\alpha_4'=\mbox{$3\over8$}\alpha_4$,
$\alpha_6'=-\mbox{$5\over16$}\alpha_6$, and so on.

\subsection{Angular distribution}

\begin{figure}[tbh]		  
\includegraphics[width=0.6\columnwidth,angle=-90]{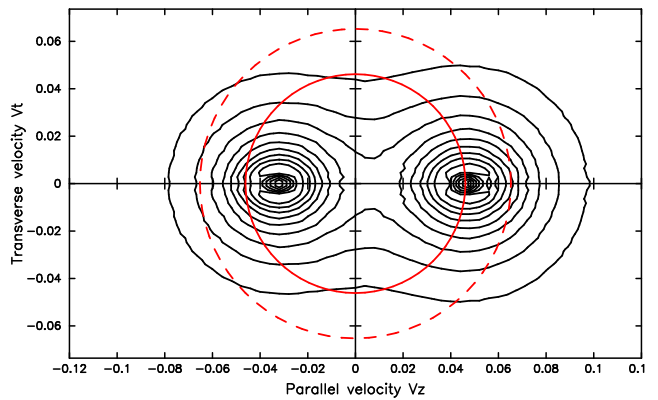}
\caption{(Color online) 
A contour plot of the velocity distribution of neutrons evaporated following
$^{235}$U($n_{\rm th}$,f) is shown.  The $z$ direction is chosen to be 
along the motion of the light primary fragment in each event.
The circles have radii corresponding to constant neutron kinetic energies
of one (solid) and two (dashed) MeV respectively.
}\label{fig4}
\end{figure}

The distributions in Eqs.~(\ref{dNdthS}) and (\ref{dNdthF}) are not 
directly observable and must be transformed to the laboratory frame,
with the associated boost velocity depending on the mass and kinetic energy 
of each fragment.
Figure~\ref{fig4} shows a contour plot of the resulting combined
velocity distribution of the neutrons from both fragments
in each event, $dN_{ n}/d^3\bld{v}$, for $^{235}$U($n_{\rm th}$,f).
Even though the distribution retains the axial symmetry of the contributions
from each fragment, it is forward-backward asymmetric 
because the light fragment moves faster and tends to evaporate more neutrons.
The circles centered at the origin
represent constant neutron kinetic energies of 1 and 2 MeV.
They make it apparent that introducing an energy threshold
will enhance the forward-backward character of the emission pattern.

\begin{figure}[tbh]   	    
\includegraphics[width=0.95\columnwidth]{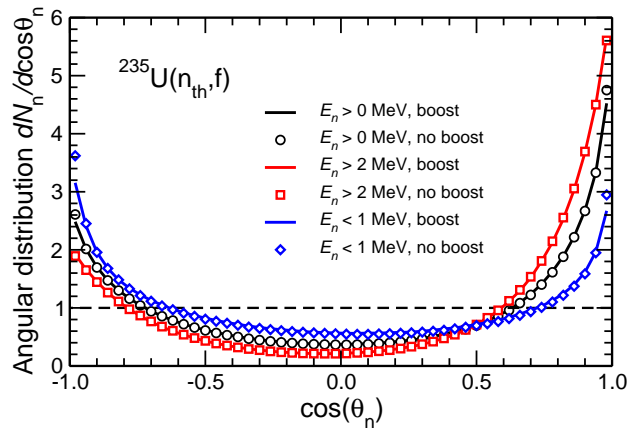}
\caption{(Color online) 
The laboratory angular distribution of the evaporated neutrons from
$^{235}$U($n_{\rm th}$,f) relative to the direction of the light primary
fragment, $dN_{n}/d\cos\theta_{n}$, 
is shown for both the standard \code\ scenario when the neutrons acquire 
a rotational boost (curves)
and a modified scenario when this boost is omitted (symbols).
In addition to considering neutrons of all energies (black),
the figure also shows the result of including either 
only neutrons with energy above 2~MeV or below 1~MeV.
}\label{fig5}
\end{figure}

The resulting angular distribution with respect to the direction
of the light fragment, $P(\cos\theta_{nL})$,
follows from the dumbbell-shaped distribution shown in Fig.~\ref{fig4}.
The boost enhances the yield in the forward direction,
$\cos\theta_{nL} > 0$, and depletes it in the backward direction,
$\cos\theta_{nL} < 0$, as shown in
Fig.~\ref{fig5} for $^{235}$U($n_{\rm th}$,f).
The black curve shows the standard result, including the rotational boost,
while the black circles show the effect of omitting that boost.
No energy cut is applied to the emitted neutrons in either scenario.
Also shown are the results for two different scenarios
when either only neutrons above 2~MeV or below 1~MeV are considered.
[As expected from Fig.~\ref{fig4},
the 2~MeV threshold enhances the relative yield in the forward direction
whereas the 1~MeV upper bound reduces the anisotropy.]
In all three scenarios, there is little visible effect of the rotation.
The largest deviations (still barely noticeable)
occur near $0^\circ$ and $180^\circ$ and for the lowest neutron energies,
which are the hardest to measure.
Thus, even though there is a clear effect of the fragment rotation
on the inherent neutron emission pattern, 
as expressed in Eqs.~(\ref{dNdthS}) and (\ref{dNdthF}),
it has a minimal influence on the observable angular distribution
$dN_{n}/d\cos\theta_{nL}$.
It thus seems unlikely that the fragment rotation can be determined
experimentally on the basis of the one-neutron distribution alone.

\subsection{Gated angular correlations}

We now investigate the sensitivity of the neutron-neutron angular correlations
to the angular momenta of the fragments.  Because there is a close correlation
between the fragment angular momentum and the total photon multiplicity, as
shown in Fig.~\ref{fig3}, this type of measurement could provide additional
information on event-averaged neutron-photon correlations beyond those measured
in Refs.~\cite{Nifenecker,Glassel,Wang} which do not provide a very clear
picture.  All three previous measurements are based on mass-averaged neutron and
photon multiplicities in different TKE bins.  Nifenecker {\it et al.}\
\cite{Nifenecker} suggested a strong positive correlation, Glassel {\it et al.}\
\cite{Glassel} saw a much weaker correlation, and Wang {\it et al.}\ \cite{Wang}
determined a more complex correlation by studying the correlation in different
mass regions.  Here we propose measuring the two-neutron angular correlations
gating on the total photon multiplicity.

\begin{figure}[tbh]
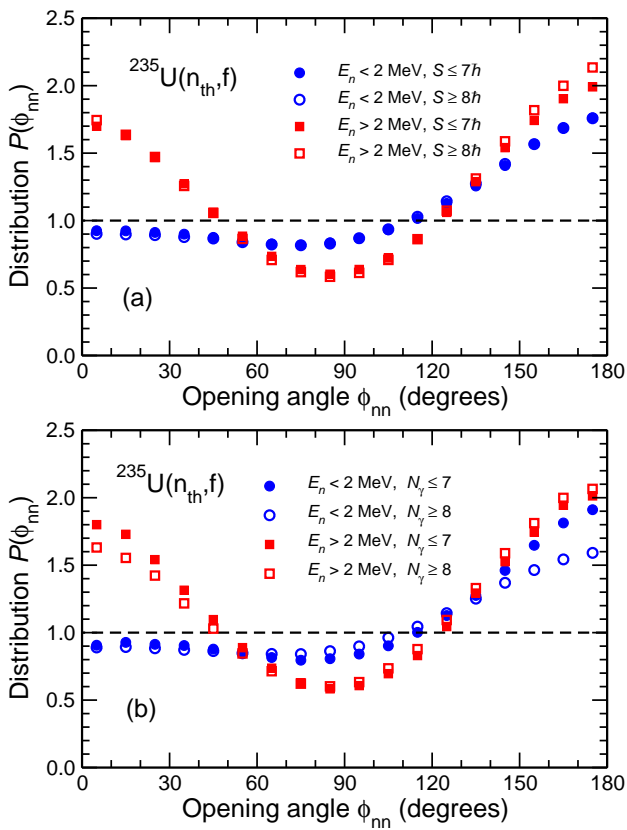

\includegraphics[width=0.95\columnwidth]{fig6a.eps}
\includegraphics[width=0.95\columnwidth]{fig6b.eps}
\caption{(Color online) 
Two-neutron angular correlations are shown for $^{235}$U($n_{\rm th}$,f), 
gating on either the total spin magnitude $S=S_L+S_H$ (a)
or the total photon multiplicity $N_\gamma$ (b).
The angular correlation functions $P(\phi_{nn})$
for either ``soft'' neutrons ($E_{n}<2$~MeV)
or  ``hard'' neutrons ($E_{n}>2$~MeV) are presented
in (a) for events with either low or high spin magnitude
($S\le7\hbar$ or $S\ge8\hbar$, respectively)
and in (b) for events with either low or high total photon multiplicity
($N_\gamma\le7$ or $N_\gamma\ge8$, respectively).
}\label{fig6}
\end{figure}

The general form of the correlation function has been discussed previously
\cite{VR-PRC90}.
It can be readily understood from the dumbbell shape of the 
neutron velocity distribution shown in Fig.~\ref{fig4}
that the distribution of the opening angle $\phi_{nn}$
is enhanced near $0^\circ$ and $180^\circ$.
For a more detailed understanding,
we note that in the case considered, $^{235}$U($n_{\rm th}$,f),
the most probable outcome (22\%) is that each fragment emits one neutron,
which contributes near $180^\circ$.
It is nearly as likely that the light fragment emits two neutrons
and the heavy fragment one, 
giving two contributions near $180^\circ$ and one near $0^\circ$ (17\% for each).
Finally, it is somewhat less likely for the heavy fragment to emit two neutrons
while the light fragment emits one, again
yielding two contributions near $180^\circ$ and one near $0^\circ$, (10\% for
each).  
The greater number of contributions to the large-angle contribution,
near $180^\circ$, results in a somewhat higher peak than at small angles,
near $0^\circ$.

Figure~\ref{fig6}(a) 
shows how the distribution of the opening angle 
between two detected neutrons, $P(\phi_{nn})$, depends on the energy
of the neutrons and the combined spin magnitudes of the two
primary fission fragments, averaged over all mass and charge partitions
as well as the TKE distribution.
Based on the combined spin magnitude, $S=S_L+S_H$,
the fission events generated by \code\ are divided into either 
``low-spin'' events ($S\leq7\,\hbar$) or ``high-spin'' events ($S\geq8\,\hbar$).
A neutron angular correlation function is extracted separately
for either ``soft'' ($E_{n}<2$~MeV) or ``hard'' ($E_{n} > 2$~MeV) neutrons.
It is evident that there is very little sensitivity to $S$,
whereas the small-angle behavior of the correlation function depends
significantly on the neutron energy.
As expected from the velocity distribution in Fig.~\ref{fig4},
the hard, energetic neutrons exhibit the expected enhancements
near $0^\circ$ and $180^\circ$,
whereas the soft neutrons do not display a small-angle peak.
  
As already discussed, although the spin magnitude $S$ is not directly
observable, the photon multiplicity may, to some degree, provide a proxy.
To illustrate this possibility,
Fig.~\ref{fig6}(b) shows how the results in Fig.~\ref{fig6}(a)
are modified when the combined spin magnitude $S$ is replaced by the 
total photon multiplicity $N_\gamma$.
Here ``photon-poor'' events having $N_\gamma\leq7$ replace low-spin events
and ``photon-rich'' events with $N_\gamma\geq8$
replace high-spin events.
There is a larger sensitivity to $N_\gamma$ than to $S$.  There is a somewhat
stronger small-angle enhancement for the photon-poor correlations with
$E_n > 2$~MeV while the large-angle peak is enhanced for low energy neutrons in
photon-poor events.

\subsection{Projected angular correlations}

A recent experimental investigation \cite{Chietera} introduced
a new analysis method for extracting a dynamical anisotropy,
the bulging of the neutron emission pattern
caused by the rotation of the evaporating fragment.
We employ \code\ to examine this idea in this section, enabling us to
assess the importance of the various effects that complicate the analysis.
Because Ref.~\cite{Chietera} studied $^{252}$Cf(sf), we focus on this case here.
However, we note that our findings apply generally.

The analysis \cite{Chietera} is based on the assumption that
the fragment spins are perpendicular to the fragment motion.
As we have already discussed, in any given event, the angular distribution
of the evaporated neutrons is (slightly) expanded in the plane 
transverse to the spin of the emitting nucleus (see Fig.~\ref{fig2}).
This plane is randomly oriented in a direction approximately perpendicular 
to the fragment velocity.
If this is indeed so, then, in each event, the transverse distribution of
the neutrons,
obtained by projecting the full three-dimensional neutron velocity distribution
onto the plane transverse to the motion of the light product nucleus,
is also distorted.
[If the contours of the full distribution are oblate spheroids
with the symmetry axis along the fragment spin
(hence perpendicular to the fragment velocity),
then the contours of the projected distribution are ellipses
with minor axes in that spin direction.]
Consequently, the distribution of the neutron-neutron opening angles,
$\phi_{nnL}$, in the transverse plane is not entirely random 
but would exhibit slight enhancements around $0^\circ$ and $180^\circ$
\cite{Chietera}.

The angular momenta of the primary fission fragments 
are determined at scission, 
at which point they are assumed to be perpendicular to the fission axis,
the line between the centers of the two fledgling fragments,
as described in Sec.~\ref{sciss}.
Subsequently, as the two fragments are being pushed apart by their
mutual Coulomb repulsion, the line connecting their centers rotates somewhat
due to the orbital motion of the dinuclear system.
Furthermore, each evaporation process changes the magnitude and direction
of both the linear and the angular momentum of the emitting nucleus.
These effects complicate the extraction of the proposed correlation signal, 
as we now discuss.

We first consider a simplified scenario in which the relative fragment motion
is purely radial, so there is no directional change of the dinuclear axis
during the Coulomb acceleration, 
and no recoils are imparted to the fragments by evaporation.
Then, to leading order, the undulating $\phi_{nnL}$ distribution 
is of the form $P(\phi_{nnL})\sim 1+c_2\cos 2\phi_{nnL}$ with $c_2>0$.
Because the amplitude $c_2$ grows approximately 
as the three-halves power of the anisotropy $A$,
it is rather challenging to extract $c_2$ for small anisotropies.
Therefore, to artificially enhance the signal for our present studies,
we have increased the \code\ ``spin temperature'' parameter $c_{S}$
from its standard value of 0.87 to 1.4.
The mean light and heavy fragment spins are then $7.3\hbar$ and $9.0\hbar$
and the angular distribution of the evaporated neutrons 
relative to the spin direction of their respective mother fragments
is then characterized by an overall dynamical anisotropy of $A\!\approx\!0.12$,
a value rather similar to that obtained by Gavron \cite{Gavron}.

If the neutrons are sampled from a common distribution with that anisotropy,
the amplitude of the angular undulation of the projected opening angle
would amount to $c_2 \approx 0.18\%$.
However, there are two distinct anisotropic distributions
in each fission event, one for each of the two fragments.
Because the fragment spins are not mutually aligned,
see Fig.~\ref{fig1}, the resulting signal is correspondingly reduced.
If the angle between the two spins were totally random,
which is very nearly the case in \code\, as discussed in Sec.~\ref{sciss},
then $c_2$ would be reduced by a factor of two.

To understand the effect of the various complications mentioned above,
we start from a simplified scenario in which the dinuclear motion
remains purely radial, as would be the case if the dinuclear complex
had no orbital motion and the linear and angular momentum recoils were absent.
In that ideal scenario, the undulation amplitude is
$c_2 \approx 0.042\%$ if all the neutrons are included in the analysis.
In an actual experiment, there is an energy threshold
below which neutrons cannot be measured.
Therefore, to conform with the experimental analysis \cite{Chietera},
we use $E_{\rm min}=0.9$ MeV in the following.
This exclusion of the softest neutrons reduces the statistics by about 
one third, while it enhances the signal somewhat, to $c_2 \approx 0.058\%$.
Generally, the \code\ simulations suggest that the effect grows
steadily stronger as a function of the threshold energy $E_{\rm min}$.

The rotation of the dinuclear axis during the separation
is typically about $2^\circ$.  The rotation 
affects the signal by just a few percent because it merely causes a
corresponding slight reorientation
of the oblate velocity distributions relative to the fragment motion.

By contrast, the effect of the spin recoils is substantial,
presumably because the evaporations change the spin direction considerably.
On average, the rotational axis of the fragment changes
its direction by about $9^\circ$ as a result of an evaporation, 
producing a corresponding tilt in the angular distribution 
of the subsequent neutron.
The \code\ simulations suggest that this effect can reduce the signal
by more than a factor of two.

Finally, we address the recoils of the linear fragment momenta 
which can significantly affect the extracted $\phi_{nnL}$ distribution.
On average, the momentum recoil from an evaporation changes the
direction of the fragment motion by only about $0.2^\circ$,
consistent with the fragment being about one hundred times 
heavier than the neutron.
Even though this directional change is very small, it nevertheless
affects the distribution of the projected opening angles to such a degree
that the undulation signal is overwhelmed.

The signal is swamped because the recoil-induced
change in the direction of fragment motion 
causes the transverse plane to be correspondingly tilted.
As a particular consequence, the projected distribution of the neutrons 
from the heavy fragment is not centered at the origin of the tilted
transverse plane but is shifted off center 
in the direction of the transverse velocity of the light fragment.
This geometric feature, which is not related to the dynamic anisotropy,
enhances the relative occurrence of small opening angles.
The resulting modulation of the distribution of the projected opening angles is,
to a good approximation, proportional to $\cos\phi_{nnL}$.

Thus the effect of the evaporation recoils 
and the effect of the dynamical anisotropy
are largely independent.  They can thus be extracted 
by performing a Fourier analysis of the distribution function,
\begin{equation}
P(\phi_{nnL})\ \sim\ 1 + c_1\cos\phi_{nnL} + c_2\cos2\phi_{nnL}\ .
\end{equation}
This approach has the additional advantage that the Fourier coefficients
can be extracted with a reasonable degree of confidence
even in the presence of large statistical errors on the individual 
values of $P(\phi_{nnL})$.
This is an important advantage 
because quite large event samples are required for the extraction of this effect
(tens of millions of events are needed in the \code\ simulations).

We finally note that the effects of the evaporation recoils 
and the rotational boosts are modified if separate analyses are made
of neutron pairs that are emitted into opposite hemispheres
(one is moving forward and the other backward, as seen in the laboratory)
and neutron pairs emitted into the same hemisphere
(both are either moving forward or backward).
The \code\ simulations provide a quantitative impression of these effects.
In the present scenario, with $c_{S}=1.4$ rather than 0.87,
\code\ yields $c_2\approx 0.040\%$.
If only pairs originating from the same hemisphere are included,
then $c_2$ is reduced to about $0.020\%$,
while it is increased to about $0.164\%$ when only pairs from opposite
hemispheres are included.
For the same cases, $c_1$ is 1.3\%, 1.2\%, and 1.6\% respectively.

\begin{figure}[tbh]	   
\includegraphics[width=0.95\columnwidth]{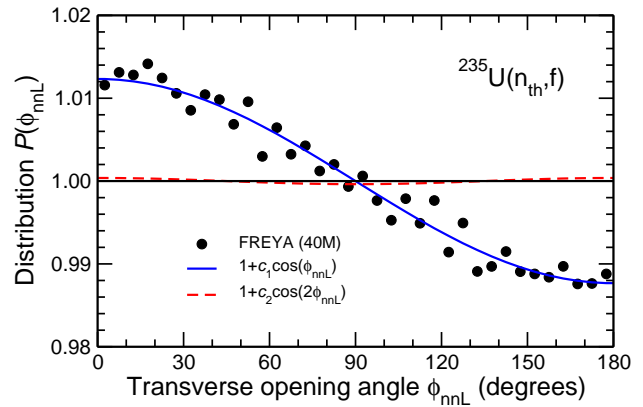}
\caption{(Color online) 
The distribution of the neutron-neutron opening angle in the plane transverse 
to the direction of the light product nucleus, $\phi_{nnL}$, 
is shown for $^{235}$U($n_{\rm th}$,f) 
as obtained from 40 million \code\ events (solid dots).
The extracted first and second Fourier components are also shown.
}\label{fig7}
\end{figure}

The results in the discussion so far are specific to $^{252}$Cf
obtained with \code\ for that case, 
but our discussion applies to other cases as well.
Figure \ref{fig7} shows the results for the case of primary interest 
in the present study, $^{235}$U($n_{\rm th}$,f).
We note that even though 40 million events were generated,
the extracted distribution exhibits 
a considerable degree of statistical fluctuation.
Nevertheless, it is possible to extract the Fourier coefficients
$c_1$ and $c_2$ with reasonable confidence and the corresponding functions
$1+c_1\cos\phi_{nnL}$ and $1+c_2\cos\phi_{nnL}$ are also shown.
The first one, resulting from the evaporation recoils, dominates,
while the undulations of the second one, reflecting the dynamical anisotropy, 
is more than an order of magnitude smaller and hardly visible.

\section{Fission from isomeric states}
\label{isomer}

In neutron-induced fission, the target nucleus is usually in its ground state
prior to the arrival of the neutron.
However, in certain environments, both in nature and in the laboratory,
there is a finite probability 
that the neutron absorption happens on an excited state of the target nucleus.
This possibility is particularly likely 
when the target nucleus has a low-lying isomeric state.
A prime example is $^{235}$U, which we shall focus on here,
whose first excited state lies at $E^*=77$~eV 
and has a half-life of around 25 minutes \cite{JETPL30,NDS40}.
 This isomeric state, $^{235{\rm m}}$U, may readily be populated in the 
astrophysical environments occurring during the $r$ process \cite{r-proc}
or in terrestrial laser-generated plasmas \cite{laser}.
The $^{235{\rm m}}$U($n_{\rm th}$,f) cross section was measured 
to be larger than the fission cross section of the ground state \cite{Deer}.

Here we seek to identify possible observable consequences of
the target nucleus being in its isomeric state rather than its ground state
when the incoming neutron arrives.
To do this, we carry out \code\ simulations for two different scenarios.
The standard scenario corresponds to the case when
the target nucleus $^{235}$U is in its ground state.
Because it has spin \mbox{$7\over2$} the resulting compound nucleus
can have angular momentum $S_0=3,4\,\hbar$. We consider $S_0=4\,\hbar$.
In the alternative scenario, the target nucleus $^{235}$U 
is in its isomeric state at 77~eV.  Because it
has spin \mbox{$1\over2$}, the resulting compound nucleus
can have $S_0=0,1\hbar$.  We consider $S_0=0$.

The potential-energy landscape for $^{236}$U 
shows a well-developed mass-asymmetric valley 
beyond the second saddle (which is asymmetric).
In addition there is a pronounced mass-symmetric valley
separated from the asymmetric valley by a down-sloping ridge.
(The topography of the fission barrier landscape is brought out
very well in Fig.\ 8 of Ref.\ \cite{IchikawaPRC86}.)
At low energy, such as occurring in $^{235}$U($n_{\rm th}$,f),
the nuclear shape evolution takes the system 
over the lowest barrier and, consequently, down the asymmetric valley.
The resulting fragment mass distribution is therefore asymmetric
and the yield at symmetry is negligible.
But as the energy is increased, it becomes ever easier
for the shape to surmount the ridge and enter the symmetric valley.
As a consequence, the mass distribution exhibits an ever more prominent
symmetric component.

A recent study using microscopic many-body level densities
to guide the Brownian shape evolution \cite{WardPRC95}
found that the ``leakage'' into the symmetric valley is sensitive
to the structure of the involved highly deformed nuclear shapes,
in particular to their pairing correlations, which may generally be larger
than the shell effects in the barrier region.
As a consequence, the symmetric yield has a delicate energy dependence
(which may even be non-monotonic).
Furthermore, because the employed combinatorial method \cite{UhrenholtNPA913}
provides the level density for different values of 
the total angular momentum $S_0$,
it was possible to also study the dependence of the symmetric yield on $S_0$
\cite{WardPRC95}.
It was found that the fragment mass distribution is generally 
rather insensitive to $S_0$ for moderate values up to $10\hbar$.  However,
the symmetric yield is significantly enhanced
for $S_0=0$ due to pairing effects.

\begin{figure}[b]  
\includegraphics[width=0.95\columnwidth]{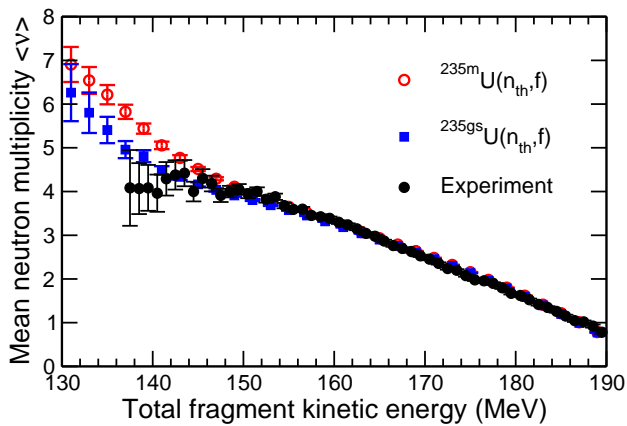}
\caption{(Color online)
The neutron multiplicity $\nu$ as a function of TKE
for $^{235}$U($n_{\rm th}$,f) in two different target scenarios:
In the first scenario (solid blue squares), 
the target is the ground state $^{235{\rm gs}}$U 
and the spin of the fissioning compound nucleus $^{236}$U is $S_0=4\,\hbar$.
In the alternative scenario (open red circles),
the target is the isomeric state $^{235{\rm m}}$U and $S_0=0$.
In both cases, 
the symbols indicate the average multiplicities at the particular TKE.
The dispersion of the multiplicity distribution at a given TKE 
is indicated by the error bars (so these are {\em not} uncertainties 
on the calculated results which are based on $4\times 10^6$ \code\ events).
The experimental data (solid dots)
are from G\"o\"ok {\em et al.}\ \protect\cite{Gook}.
}\label{fig8}
\end{figure}

This finding is of particular interest to our present study,
because the isomeric state in $^{235}$U leads to compound spins of
$S_0=0,1\hbar$ with about equal probability,
whereas the ground state leads to $S_0=3,4\hbar$.
In the latter case, for which extensive experimental data exist,
the symmetric yield is very small, 
while the results reported in Ref.~\cite{WardPRC95} suggest that for $S_0=0$
the symmetric mass yield is about $5\%$ of the peak yield.
We wish to explore the consequences of such a possible enhancement.

We have therefore constructed a mass distribution 
with a suitably enhanced symmetric yield to use as \code\ input.
This is relatively easily done, because the usual input mass distribution
is represented as a sum of a dominant asymmetric contribution
and a small symmetric component \cite{Brosa}.
We have increased the relative weight of the symmetric term to ensure
$Y({\rm symm})/Y({\rm peak})=0.05$
and we use this modified mass distribution
when simulating the alternative scenario
where the target nucleus is in its isomeric state.  
We have left all other \code\ inputs unchanged, 
including the input TKE distribution TKE$(A_H)$.

It is important to note that the two fission modes,
asymmetric and symmetric, have other distinct characteristic features
apart from the difference in their mass splits.
Of particular relevance is the fact that the scission shapes
of the symmetric mode tend to be significantly more elongated
than those of the asymmetric mode \cite{WardPRC95,Albertsson2020}.
Thus, in the symmetric mode the centers of the proto-fragments 
are further apart than in the asymmetric mode
and the potential energy at scission is correspondingly lower.
This results in larger excitation energies at scission
and smaller fragment kinetic energies.
Furthermore, the additional fragment excitation
gained from the shape relaxation of the distorted proto-fragments
\cite{Albertsson2020} is also larger than in the asymmetric mode.
The resulting higher excitation of the symmetric-mode primary fragments 
will in turn cause more neutrons to be evaporated.
Consequently, one should expect the two scenarios 
to exhibit different relations between
the number of neutrons evaporated and the total fragment kinetic energy
in the region of low TKE.

\begin{table}
\begin{tabular}{c|ccccc}
{\rm Case} & 
     	   &${\cal M}_1$ &  ${\cal M}_2$ & ${\cal M}_3$ & ${\cal M}_4$\\
\hline\\[-2.5ex]
$^{235{\rm gs}}$U($n_{\rm th}$,f), $S_0\!=\!4\hbar$~ & 
	   & 2.39526\,\, & 4.53167\,\, & 6.46183\,\, & 6.59926\\
$^{235{\rm m}}$U($n_{\rm th}$,f), $S_0 = 0$~ & 
	   & 2.43591 & 4.75387 & 7.23551 & 8.62523\\
\end{tabular}
\caption{Factorial moments of the neutron multiplicity distribution,
${\cal M}_n=\langle\nu(\nu-1)\dots(\nu-n+1)\rangle$, for thermal fission 
using the ground state or the isomeric state of $^{235}$U. 
}\label{facmom}
\end{table}

\begin{figure}[b]	
\includegraphics[width=0.95\columnwidth]{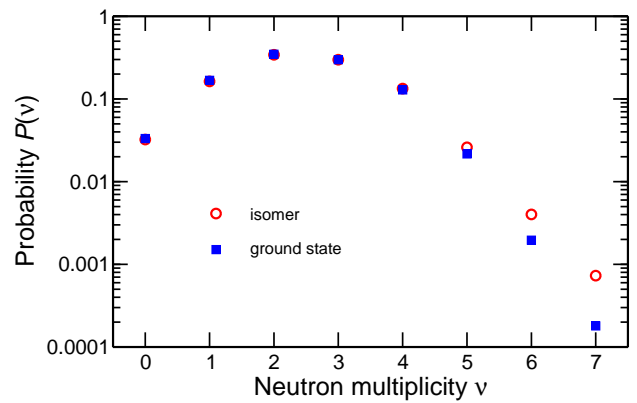}
\caption{(Color online)
The neutron multiplicity distribution $P(\nu)$ for 
the two cases of $^{235}$U($n_{\rm th}$,f) shown in Fig.\ \ref{fig8}:
$^{235{\rm gs}}$U (solid blue squares) and $^{235{\rm m}}$U (open red circles),
shown on a logarithmic scale in order to better bring out
the enhancement of high-multiplicity events.
}\label{fig9}
\end{figure}

This is indeed the case, as illustrated in Fig.~\ref{fig8} 
which shows the multiplicity of evaporated neutrons as a function of TKE.  The
results for the two cases are effectively identical for ${\rm TKE}\geq150$~MeV
but begin to separate for lower values where the higher probability for fission
from the symmetric mode in the isomeric state tends to yield higher neutron
multiplicities at low TKE.  The data from Ref.~\cite{Gook}, also shown on
Fig.~\ref{fig8}, are consistent with both calculations above 150~MeV but tend
to agree more with the ground state calculation, within the increasing
statistical uncertainties, until TKE$\approx 140$~MeV where the data set ends.
If data could be taken to still lower TKE, it might be possible to better
distinguish between the two scenarios.  Also, if isomeric-state targets could be
fashioned and used to obtain a sufficiently significant data set, it might be
feasible to measure a difference between the two scenarios.  However, this
would require a large number of isomeric targets to obtain enough low TKE data
to observe statistical differences.

On the theoretical side, these results
were obtained only by an ad hoc modification of the mass yields, $Y(A)$, to
enhance the yield at symmetry, consistent with the results of
Ref.~\cite{WardPRC95}.  An improved calculation of the shape evolution for
different values of $S_0$ would be required to obtain a more precise $Y(A)$
distribution to use in \code.  

The enhanced neutron multiplicity at low TKE from an isomeric target should 
also manifest itself in an overall larger average neutron multiplicity
and this is indeed the case.
The neutron multiplicity distribution can be conveniently
characterized by its factorial moments, 
${\cal M}_n=\langle\nu(\nu-1)\dots(\nu-n+1)\rangle$.
As seen in Table~\ref{facmom}, the average neutron multiplicity,
the first factorial moment ${\cal M}_1$, is increased by 1.7\%.  
As evident in Fig.~\ref{fig8}, the increase in the
total multiplicity comes from the low TKE events 
which yield the highest values of $\nu$.  
The next three moments, ${\cal M}_2 - {\cal M}_4$, are also shown in
Table~\ref{facmom}.  It is clear that fission from the isomeric state enhances
the higher multiplicity moments most.  This can also be observed graphically
in Fig.~\ref{fig9}, presented on a logarithmic scale to more easily distinguish
the high multiplicity behavior.

\section{Concluding remarks}
\label{summ}

We have studied the role of angular momentum in the fission process to search
for evidence of any quantitative effect it might have on fission observables.
We employed \code\ in this study because it obeys all conservation laws
throughout each step of a fission event, from scission through prompt neutron
and photon evaporation.  All of these analyses were carried out without
changing \code\ inputs, unless
otherwise noted.  We have previously studied how changes in the spin temperature
parameter, $c_S$, modifies the fragment rotational energy and thus affects
photon observables, see Ref.~\cite{VR-PRC96} for details.

We have shown that, even if the initial compound nucleus is prepared 
with a definite angular momentum,
which endows the fragments with correspondingly aligned average spins,
the spin fluctuations acquired at scission ensure that there is little 
correlation between the resulting fragment spins 
and that of the compound nucleus.
Furthermore, the spins of the two fragments are also essentially uncorrelated.

We showed that the total photon multiplicity is related to the combined
magnitude of the two fragment spins, especially for $S_L + S_H > 5\hbar$,
see Fig.~\ref{fig3}.
We found that this effect is almost independent of the incident neutron energy,
perhaps because, in \code, 
neutrons are emitted as long as energetically possible.

We have also studied neutron observables and found that neutron emission
from a rotating fragment results in a oblate emission pattern,
as first discussed in Ref.~\cite{RV-PRC89}.
The resulting dynamical anisotropy increases with the fragment spin
but it is hardly sensitive to the spin of the fissioning nucleus.
Because the dynamical anisotropy is relatively small ($\sim10\%$),
it has hardly any observable influence on the neutron angular distribution
with respect to the direction of the light fragment in the laboratory,
see Fig.~\ref{fig5}.

We have particularly focused on correlation observables.  We discussed using
neutron-neutron correlations gated on photon multiplicity as a proxy for gating
on fragment angular momentum, see Fig.~\ref{fig6}.
While we found a weak dependence of the angular
correlation on photon multiplicity, we also showed that making distinction
between low and high energy neutrons has a stronger impact on the correlation
function.  Finally, we discussed the
projected neutron-neutron angular correlations
proposed in Ref.~\cite{Chietera} and found that even though
the signal of the dynamical anisotropy is weak 
and is overwhelmed by the effect of evaporation recoils 
on the linear and angular momenta of the emitting fragments,
it may be extracted by Fourier analysis, see Fig.~\ref{fig7}.

In a separate analysis, we discussed possible observable effects of
neutron-induced fission on the isomeric state of $^{235}$U
instead of a $^{235}$U target in its ground state.
For this, we employed a modified yield function $Y(A)$ to model the enhanced 
symmetric yield from a spin-zero $^{236}$U compound nucleus 
obtained in Ref.\ \cite{WardPRC95}
and found that this results in higher neutron multiplicities at low TKE, 
a potentially observable effect that could distinguish fission 
from the isomer relative to the ground state, see Fig.~\ref{fig8}.
[It would obviously be very interesting to experimentally test 
the enhancement of the symmetric yield for spin zero predicted 
in Ref.\ \cite{WardPRC95}.]

In general, we found that angular momentum effects are subtle 
and are generally insensitive to the spin of the initial state.  
While we have primarily focused
on $^{235}$U($n$,f), we have found similar effects for other isotopes.  The
exception is fission from the low-lying uranium isomeric state where 
the predicted increase in the symmetric yield 
could have observable consequences if enough target
material were available to accumulate sufficient statistics at low TKE.

\section*{Acknowledgments}
This work was supported by the Office of Nuclear Physics in the U.S.
Department of Energy under Contracts DE-AC02-05CH11231 (JR) and
DE-AC52-07NA27344 (RV) and was supported by the LLNL-LDRD Program under Project
No.~20-ERD-031 (RV).

\end{document}